\documentclass[12pt]{article}

\usepackage[utf8]{inputenc}

\usepackage[margin=1.25in]{geometry}

\usepackage{setspace}
\usepackage{natbib}       
\usepackage{graphicx}     
\usepackage{booktabs}     
\usepackage{amsmath}      
\usepackage{authblk}      
\usepackage{hyperref}     
\usepackage{url}          
\usepackage{algorithm}    
\usepackage{algorithmic}

\title{Randomized Recruitment Driven Sampling}
\author[1]{Adam Visokay}
\author[2]{Laura Boudreau}
\author[3]{Rachel Heath}
\author[1,4]{Tyler H. McCormick}
\affil[1]{Department of Sociology, University of Washington}
\affil[2]{Columbia Business School, Columbia University, \& CEPR}
\affil[3]{Department of Economics, University of Washington}
\affil[4]{Department of Statistics, University of Washington}
\date{\today}
\begin{document}

\maketitle

\linespread{1}{
\begin{abstract}
    \noindent Surveys are critical inputs for research and policy, yet, enumerating a sampling frame is logistically infeasible or financially nonviable in many circumstances, such as during pandemics, natural disasters, or armed conflict. Respondent Driven Sampling (RDS) does not require a sampling frame, yet non-random peer recruitment often introduces substantial bias, particularly under high homophily. We introduce and evaluate Randomized Recruitment Driven Sampling (RRDS), a cellphone-based adaptation of RDS that incorporates researcher-controlled randomization into each recruitment wave. While standard RDS is necessary for stigmatized groups where network transparency is infeasible, RRDS is designed for low-stigma populations that become difficult to access due to logistical barriers. In these contexts, RRDS enforces the random recruitment assumption that traditional RDS relies upon but rarely achieves. Through simulation and an experiment surveying Bangladeshi garment workers during the COVID-19 pandemic, we demonstrate that RRDS produces less biased estimates and improved confidence interval coverage compared to traditional RDS. RRDS offers a scalable, remote-compatible alternative for studying low-stigma groups in challenging contexts where large-scale probability sampling is unsafe or infeasible.
\end{abstract}
}

\newpage
\section{Introduction}

When studying accessible populations not subject to stigma, researchers typically use probability sampling to ensure representative samples. This approach relies on comprehensive lists of individuals within the target population—such as voter registration records, customer databases, census registries, or administrative files. These population inventories, known as sampling frames, enable researchers to select representative samples through probability-based survey sampling \citep{kish1965surveysampling, kalton1983surveysampling}. Such data serve as the foundation for empirical work across sociology, political science, economics, and public health. Alternatively, researchers can approximate random sampling by directly sampling from a population, such as selecting every Nth residence in a geographic area \citep{groves2009survey, pearson2015using}. This approach allows researchers to weight their samples with appropriate inclusion probabilities, ensuring representativeness and enabling valid population-level inferences.

While probability samples are the gold standard for statistical inference, they are not always feasible to collect. Complete population lists are often prohibitively expensive to obtain and maintain.  Global disruptions including the COVID-19 pandemic, floods, fires, other natural disasters and armed conflict have created further barriers to data collection for populations that would normally be accessible through traditional methods.  Unlike populations that are inherently difficult to study due to stigma or marginalization (e.g. the unhoused, injection drug users, sex workers), these groups may become inaccessible due to external circumstances such as mobility restrictions or heightened health concerns. These contextual constraints have accelerated the need for innovative, remote sampling methods that maintain methodological rigor while adapting to situational rather than population-based barriers to access. 

One common strategy to reach respondents in the absence of a population list or sampling frame is to use respondents' social networks to access members of the population. Respondent Driven Sampling (RDS), for example, is a network-based sampling procedure that leverages social structures to recruit participants in the absence of a sampling frame. Initially proposed by \cite{heckathorn1997}, RDS has become a standard approach for accessing hard-to-reach (typically because of stigma or exclusion from formal institutions/resources) populations worldwide.  In a classic RDS set-up, respondents are given incentives both to fill out surveys themselves and for each referral they make who is successfully surveyed.  Despite being widely adopted, the principal challenge associated with RDS data lies in the inherent difficulty of quantifying recruitment bias: given that the respondent has incentive to refer people who are most likely to successfully complete a survey, referred individuals may differ from the population as a whole.\footnote{Indeed, in labor market settings, people give different referrals when given an incentive based on the job performance of their referral compared to if they do not \citep{beaman2012gets}.} When recruitment bias exists, traditional RDS estimators can produce biased results, particularly in populations with strong within-group ties or low likelihoods of completing surveys, or when sampling a large fraction of the total population \citep{gile2011improved}. It is also difficult to validate these approaches, as population parameters for hard-to-reach populations are by definition unknown. These limitations are well-documented in the RDS literature \citep{gile2010respondent, salganik2004sampling, volz2008probability}, yet viable alternatives for accessing these populations have remained elusive. 

We introduce Randomized Recruitment Driven Sampling (RRDS), a variant of RDS designed specifically for low-stigma populations that have become difficult to access due to external circumstances (e.g. natural disasters, armed conflict). Where in RDS, each respondent refers a few contacts (and is given an incentive if those referrals complete a survey), RRDS elicits contact information from a large share of the respondent's network (with incentive given for each referral made, regardless of whether the referral results in a completed survey\footnote{In theory, the provision of payment for each person referred could create incentive for respondents to make up names/phone numbers to increase their payment.  The payment per phone number we gave -- 10 taka, or 0.12 USD in May 2021 -- was relatively modest, and consistent with minimal false reports, only 4.7\% of calls answered were answered by those who did not know the name of the contact listed.  Nonetheless, we would recommend that other researchers conducting RRDS monitor for signs of false reports.}) and then samples randomly from it.  While this approach  would likely be infeasible in the stigmatized populations for which RDS was originally developed, when implemented in low-stigma populations it directly addresses the well-documented problem that respondents do not recruit randomly from their networks. 

This method has several distinctive features: (1) it reduces physical contact requirements, making it particularly suitable for disaster settings and public health emergencies; (2) it works naturally with increasingly ubiquitous mobile phone technology; (3) it introduces randomization procedures that help quantify and mitigate recruitment bias; (4) it provides a scalable platform for reaching hard to reach populations while maintaining rigorous statistical properties; and (5) because we often do have access to baseline population parameters for these groups, we can successfully evaluate network-based sampling performance in this setting. Drawing from both a simulation study and an experiment where we survey Bangladeshi garment workers during the COVID-19 pandemic, we demonstrate the effectiveness of this randomized recruitment approach.

The remainder of this paper is organized as follows: in Section \ref{sec:rds} we review the theoretical foundations and limitations of traditional RDS. In Section \ref{sec:rrds} we detail our proposed randomized recruitment strategy (RRDS) and in Section \ref{sec:sim} we demonstrate performance with a simulation study. Section \ref{sec:data} describes the treatment and control arms of the recruitment experiment and Section \ref{sec:results} details the results of the analysis. Finally, we discuss implications for survey research methodology and directions for future refinement of the RRDS approach in Section \ref{sec:discussion}.

\section{Related Work on Network-Based Sampling Methods}
\label{sec:rds}

Network-based sampling has a long history in social science, beginning with snowball sampling \citep{goodman1961snowball, biernacki1981snowball, heckathorn1997}. In this approach, an initial set of contacts recruits further participants from their own networks, who in turn recruit others iteratively. Snowball sampling is highly effective at reaching hidden or hard-to-reach populations where complete enumeration (like a register) or a probabilistic sampling frame (like a census) is not feasible. However, it carries a well-known limitation: biased samples \citep{biernacki1981snowball, goodman2011comment}. Because each wave of recruits is selected by the preceding wave rather than by the researcher, the resulting sample lacks well-defined inclusion probabilities, making traditional population-level inferences invalid.

To address this gap, Respondent-Driven Sampling (RDS) builds on traditional chain-referral by combining peer recruitment with a formal statistical framework grounded in Markov-chain theory \citep{heckathorn1997, salganik2004sampling, goel2009rdsmcmc}. By eliciting information about each respondent's network degree, RDS calculates inclusion probability weights to convert a non-probability snowball sample into a tool for population-level inference. This methodology has become a popular paradigm for studying vulnerable or stigmatized populations, such as the unhoused \citep{almquist2025innovating, wesson2025novel}, injection drug users \citep{robinson2006recruiting}, sex workers \citep{szwarcwald2011analysis, yamanis2013empirical}, and victims of human trafficking \citep{Kunkeetal2023}. Recent adaptations have expanded its practical reach through alternative implementation channels, beginning with web-based RDS \citep{wejnert2008web} and including phone-based surveys \citep{inghels2021telephonerecruitment}, internet- and cell-phone-based remote sampling \citep{pham2023remotesampling}, smartphone data collection \citep{Sosenko2022}, and SMS/WhatsApp web interfaces \citep{McGowan2023}.

Rather than relying purely on passive peer referral, some work attempts to change the structural recruitment process itself. \cite{Mouw2012NSM} propose Network Sampling with Memory (NSM), which collects network data from respondents directly and uses it to adaptively sample bridge nodes to achieve faster network coverage. NSM requires maintaining a global memory of all contacts elicited across every recruitment chain to dynamically update sampling weights. While this method has been used successfully in the field \citep{merli2015migration, merli2022demography}, the required coordination infrastructure (typically a live database updated between waves) can be difficult to sustain in the resource constrained field settings that motivate our work, such as disaster zones, conflict areas, or pandemic conditions.  The strategy is also fundamentally different from ours in that it focuses on global coordination, while our approach uses a local randomization and leverages our ability to directly influence recruitment because groups are not highly stigmatized.

Because our proposed method builds directly on the paradigm established by RDS, we outline its core operational mechanics and key theoretical assumptions. Operationally, an RDS study begins with a convenience sample of individuals within the target population, known as ``seeds.'' Seeds receive a primary incentive for completing the survey and a secondary incentive to recruit a fixed number of their peers. This cycle repeats across successive waves until a target sample size is achieved. Theoretically, standard RDS estimation rests on six critical assumptions \citep{heckathorn1997, salganik2004sampling, volz2008probability, gile2010respondent, gile2014diagnostics}:

\begin{enumerate}
    \item \textbf{Respondents recruit randomly from their personal networks}: Participants are assumed to select recruits uniformly at random from their eligible network connections. This allows researchers to model recruitment as a random walk on a social network.
    
    \item \textbf{Respondents accurately report their personal network size}: Estimators require participants to report their network degree ($d_i$) to calculate inverse-probability weights. 
    
    \textbf{Volz-Heckathorn (VH) Estimator}: Grounded in these first two assumptions, the standard VH estimator \citep{volz2008probability} for the population mean $\mu$ of an attribute $x$ weights individuals by their inverse degree to avoid over-representing high-degree nodes:
    \begin{equation}
    \label{eq: VH}
        \hat{\mu}_{\text{VH}} = \frac{\sum_{i=1}^{n} \frac{x_i}{d_i}}{\sum_{i=1}^{n} \frac{1}{d_i}}
    \end{equation}
    
    \item \textbf{Sampling occurs with replacement}: Standard theory assumes an individual could theoretically be recruited multiple times. In practice, this is routinely violated as respondents do not complete the final survey more than once, leading to estimation challenges when the sampling fraction is large \citep{gile2011improved}. 
    
    \item \textbf{The target population forms a single connected component}: All members of the target population must be connected directly or indirectly through the social network. If the population instead comprises multiple components, seeds are required in every component for full coverage, and there is no principled way to weight estimates across components.
    
    \item \textbf{Recruitment reaches the full network}: Chains must progress through enough waves (typically six or more) to reach a sampling equilibrium that overcomes the non-random selection of the initial seeds. This assumes participants are sufficiently motivated to refer contacts \citep{semaan2009ethical, mccreesh2013community}; otherwise, chains die out. Premature chain death reduces not only sample size but also introduces bias, since the realized sample over-represents the network neighborhoods of the chains that survive.
    
    \item \textbf{Reciprocity of relationships}: Network ties are assumed to be reciprocal (i.e., if Person A reports a tie to Person B, Person B would report a tie to Person A).
\end{enumerate}

In practice, the foundational assumption of random recruitment is routinely violated due to homophily and convenience selection: respondents preferentially refer peers who are demographically or socially similar to themselves rather than selecting randomly \citep{liu2012assessment, mcreesh2012evaluation, li2018overlooked, zeng2019empirical, buchanan2022testing}. Notably, even official implementation guidance does not, and cannot, operationalize truly random recruitment: widely used field manuals instruct respondents to select recruits via heuristics (for example, encouraging recruiters to choose peers who do not know one another), which constitutes purposive rather than random selection \citep{who2013rds}. The gap between the theoretical assumption and any respondent-executed instruction motivates transferring the selection step to the researcher.

To account for these violations, a substantial line of literature has focused on developing estimation procedures that relax or correct for the random recruitment assumption after data collection. \citet{heckathorn2002rds2} proposed an estimator that derives population estimates from observed cross-group recruitment patterns rather than assuming recruitment is random. More recently, \citet{lu2013linked} and \citet{beaudry2020correcting} introduce estimators that leverage ego-network data to adjust for non-random recruitment, which simulations suggest can drastically reduce bias under realistic field conditions \citep{verdery2015respondent}. In parallel, to handle the variance underestimation caused by the dependencies within recruitment chains, \citet{baraff2016estimating} and \citet{Greenetal2020} introduced the multilevel tree bootstrap method (\texttt{RDStreeboot}), which resamples hierarchically within the recruitment tree structure to preserve the underlying dependency framework. Finally, and most closely related to our design, \citet{falorsi2023unbiased} and \citet{alleva2025building} develop a formal estimation framework for network sampling designs in which the researcher, rather than the respondent, selects recruits from elicited contact lists. Our study is complementary to that theoretical work: we provide a field implementation and an empirical evaluation of such a design against traditional RDS.

Ultimately, existing adjustments for homophily operate almost exclusively on the estimation side, attempting to model and correct for a bias that has already occurred in the field. In contrast, our proposed Randomized Recruitment Driven Sampling (RRDS) method operates on the design side. Like NSM, RRDS directly alters the recruitment strategy itself to prevent preferential recruitment from arising in the first place. Crucially, however, because RRDS is designed for settings where population members are difficult to access for external reasons rather than stigma, it can do so by intervening locally at each selection wave, bypassing the logistical database infrastructure NSM requires to influence the global direction of recruitment. Because RRDS modifies the recruitment mechanism rather than the downstream mathematics, it remains fully complementary to, and can be paired with, the post-hoc estimators and bootstrap procedures detailed above. An important open question, which we return to in the Discussion, is how design-side randomization compares to estimation-side corrections such as those of \citet{heckathorn2002rds2} and \citet{beaudry2020correcting}; because RRDS elicits ego-network lists by construction, the two approaches can in principle be combined, with randomization reducing the burden placed on post-hoc modeling assumptions.

\section{Randomized Recruitment Driven Sampling}
\label{sec:rrds}

Traditional RDS was developed for stigmatized or HTR populations—such as injection drug users or sex workers -- where respondents cannot reasonably be asked to disclose other known members and practitioners need to leverage the connectedness of the HTR group in order to access additional group members. In such settings, respondents provide only a few referrals, and the researcher has no way to enforce random selection among network ties. However, when populations become hard-to-reach due to logistical constraints rather than stigma (e.g., natural disasters, armed conflict, pandemics), a different approach becomes feasible. Randomized Recruitment Driven Sampling (RRDS) seeks to elicit each respondent's complete network and introduces researcher-controlled random selection of recruits. This design is possible precisely because stigma is not the barrier to access: respondents in low-stigma populations can provide comprehensive contact lists without fear of exposing themselves or their peers. Further, implementation for cell-phone based networks can be done entirely remotely, which allows surveying when in-person recruitment is unfeasible or prohibitively expensive. While RRDS could theoretically be deployed across other digital mediums -- such as social media platforms or email -- cell-phone networks are particularly well-suited for our motivating settings. And while traditional RDS could also be implemented remotely in these settings, the critical assumption that respondents recruit randomly from their networks is well-documented to fail in practice due to homophily and convenience selection. RRDS addresses this problem directly by removing recruitment decisions from respondents entirely. Like phone-based implementations of traditional RDS, RRDS can be conducted remotely, but its advantage lies in using contact lists as a sampling frame for uniform random sampling as recruitment. 

\noindent We formalize RRDS as a two-stage recruitment procedure. For each respondent $i$:

\begin{enumerate}
    \item \textbf{Nomination stage:} Respondent $i$ produces a list $L_i$ of size $m_i$ from their eligible network neighborhood $N_i$ of size $k_i$ (the respondent's degree).
    
    \item \textbf{Recruitment stage:} The researcher samples $s_i$ individuals uniformly at random from $L_i$.
\end{enumerate}

\noindent This two-stage structure makes explicit the sources of potential bias and the assumptions required for valid inference with RDS estimators. The validity of RRDS depends critically on the relationship between $L_i$ and $N_i$. We distinguish three such scenarios:

\begin{enumerate}
    \item \textit{Exhaustive nomination} ($L_i = N_i$, so $m_i = k_i$): The respondent reports their complete eligible network, and the recruitment stage samples uniformly from the full neighborhood.
    
    \item \textit{Approximately-exhaustive nomination}: When $L_i$ is elicited from call logs or similar records, we assume these provide an approximately complete record of active network ties, so $L_i \approx N_i$.
    
    \item \textit{Selective nomination}: If $L_i \subset N_i$ with $m_i < k_i$, we require that the selection of $L_i$ from $N_i$ is either approximately random or follows a known missingness structure that can be adjusted for in estimation.
\end{enumerate}

By taking more control of the recruitment stage, RRDS more closely satisfies a core theoretical requirement of RDS estimators. Standard RDS relies on modeling the sampling process as a Markov chain on the underlying social network graph. For this approximation to hold, recruitment must represent a random walk, requiring that respondents select peers uniformly at random from their network neighborhood. In traditional RDS, this assumption routinely fails because respondents exhibit differential recruitment and non-random selection based on convenience or tie strength. By enforcing uniform random sampling from the elicited list $L_i$, RRDS forces the empirical sampling process to more closely approximate the theoretical Markov chain behavior, satisfying the random recruitment assumption.

Furthermore, this enforced random recruitment provides a theoretical justification for applying the tree bootstrap for variance estimation. In standard RDS, standard error estimation is complicated by the dependence between recruiters and recruitees driven by network homophily. The tree bootstrap resamples the observed recruitment chains to preserve this dependence structure. In traditional RDS, the observed tree reflects a conflation of underlying network homophily and the respondent's own recruitment biases. Under RRDS with exhaustive nomination, respondent-driven selection bias is removed by design, so the dependence structure captured by the tree bootstrap reflects the network's structural homophily together with the known randomized selection mechanism, rather than an unknown mixture of structure and recruiter behavior. We therefore expect the tree bootstrap to characterize sampling variation under RRDS at least as faithfully as under traditional RDS, though we emphasize that this is a heuristic argument; developing variance estimators tailored to randomized recruitment remains an open problem (Section~\ref{sec:discussion}).

Beginning with initial seeds as in traditional RDS, the researcher randomly selects $s_i$ contacts from each respondent's list, repeating across waves until reaching the desired sample size. Because RRDS requires respondents to disclose their full network of eligible contacts, it may not be appropriate for the stigmatized populations where RDS has traditionally been applied. Asking injection drug users or sex workers to provide complete lists of their peers would raise serious ethical concerns and likely yield incomplete or inaccurate responses. RRDS is therefore best suited to populations that face logistical rather than social barriers to access. In such settings, traditional RDS remains a viable alternative—but one that relies on an assumption (random peer recruitment) that is consistently violated in practice. RRDS offers a way to satisfy this assumption by design, at the cost of requiring more extensive network disclosure from respondents.

\section{Simulation Study}
\label{sec:sim}

To evaluate the effectiveness of RRDS, we conducted a simulation study using a synthetic social network with known population parameters. This controlled setting allows direct comparison of how RDS and RRDS converge to truth from a deliberately biased starting point.

\paragraph{Network.}
We generated a synthetic social network of 10,000 nodes. Node attributes were drawn from known population distributions: age was normally distributed with mean 41.5 years and standard deviation 10, bounded to the range $[18, 65]$; gender was binary with 70\% female. Network edges were generated using a modified random graph model targeting a mean degree of 10 connections per node, consistent with empirical estimates of ego-network size in general adult populations \citep{dunbar1995social}. 

\paragraph{Two types of homophily.}
A key conceptual distinction motivates the simulation design. Homophily can arise from two separate sources that must not be conflated:

\begin{itemize}
  \item \textbf{Network homophily} ($\alpha_{\text{net}}$): the tendency of the underlying social network to connect demographically similar individuals. This is a property of the network itself. In our simulation, $\alpha_{\text{net}} = 0.40$, meaning 40\% of edges are formed preferentially between same-gender nodes (weighted toward age similarity), while the remaining 60\% are formed at random. Randomizing recruitment does \textit{not} address structural homophily: the contact list that any participant enumerates is already shaped by who they know, which in turn reflects the network's structure.

  \item \textbf{Preferential recruitment} ($\alpha_{\text{rec}}$): the tendency of RDS participants to actively recruit contacts similar to themselves from among their known neighbors. In our simulation, $\alpha_{\text{rec}} = 0.90$, meaning that in a standard RDS chain, each participant recruits a same-gender (and age-similar) contact with probability 0.90, and recruits at random with probability 0.10. This behavioral bias is precisely what RRDS is designed to eliminate. By requiring participants to randomly sample from their enumerated contact list, RRDS removes preferential selection at the recruitment step regardless of the network's structural composition.
\end{itemize}

Separating these two parameters makes clear that RRDS corrects for $\alpha_{\text{rec}}$ but not $\alpha_{\text{net}}$. Sampling from a homophilic network will always yield a sample whose composition reflects that underlying structure to some degree; RRDS ensures that the \textit{additional} bias introduced by systematic preferential recruitment does not compound this.

\paragraph{Seeds and stopping rule.}
All simulations were initialized with exactly 10 biased seeds -- young men under age 22 -- representing a ``worst-case" starting condition: seed mean age approximately 20 years and 0\% female, versus known population means of 41.5 years and 70\% female. Both RDS and RRDS began from the same seeds to isolate the effect of the recruitment mechanism. Each participant could recruit up to three contacts per wave. Recruitment continued until the \textbf{cumulative sample reached $n = 3000$}, at which point chains were stopped regardless of wave number. This stopping rule reflects the budget-constrained reality of field studies, where data collection ends when a pre-specified sample size is reached rather than when the network is exhausted. It also ensures that any observed differences between methods reflect performance within a practically meaningful sample size rather than eventual asymptotic equivalence.

\paragraph{Scenarios.}
Despite the fact that random recruitment is a core RDS assumption, we know that in practice it is often violated. Therefore, the baseline we use for comparison is RDS \textit{with} preferential recruitment ($\alpha_{\text{rec}} = 0.90$). Against this we compare the performance of Randomized Recruitment and RRDS under conditions of partial non-response and incomplete contact enumeration, as real-world implementation will face both challenges. These robustness scenarios are described and presented in Appendix~\ref{app:robustness} but do not undermine the substantive conclusion.

\begin{figure}[!ht]
  \centering
  \includegraphics[width=0.48\textwidth]{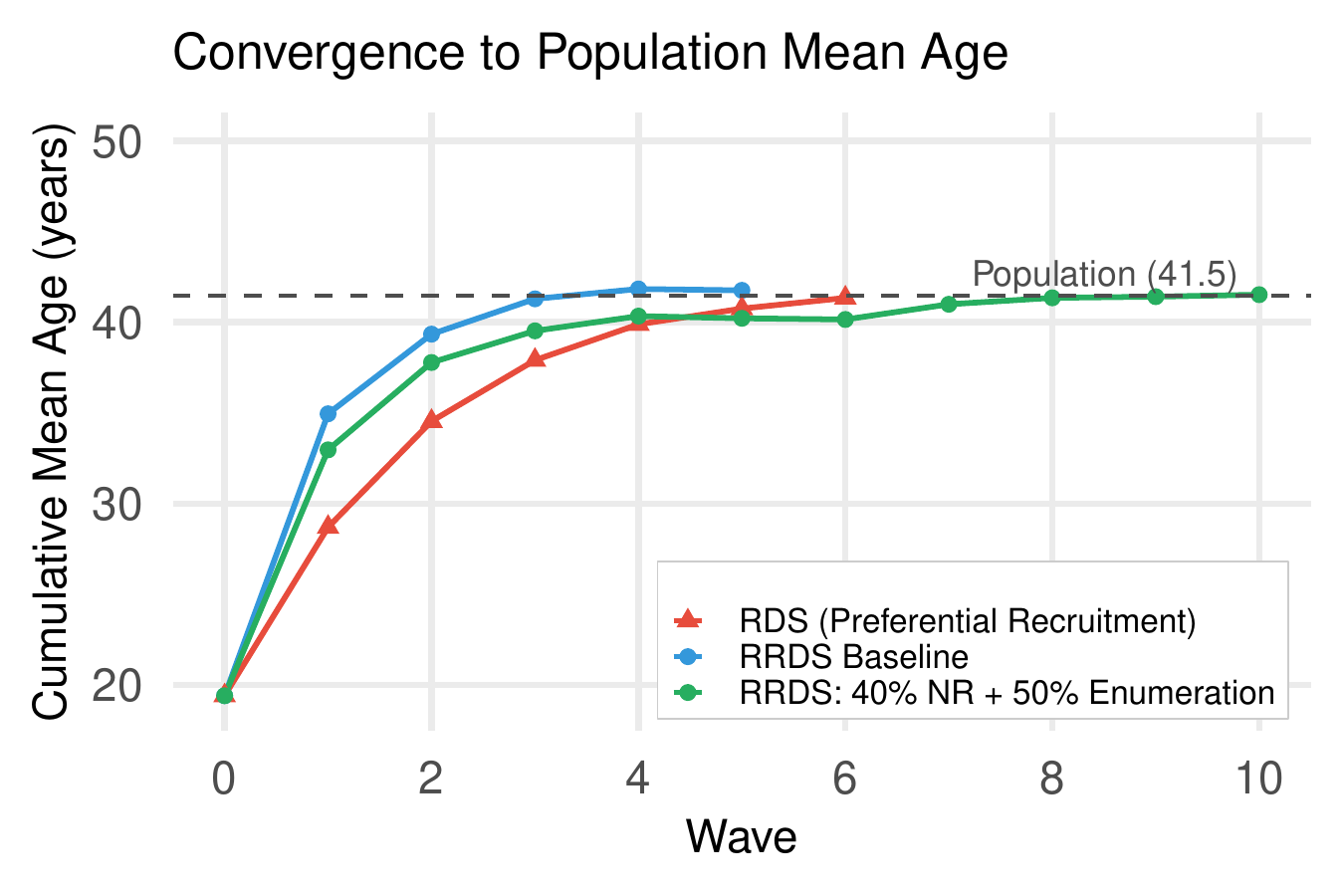}
  \includegraphics[width=0.48\textwidth]{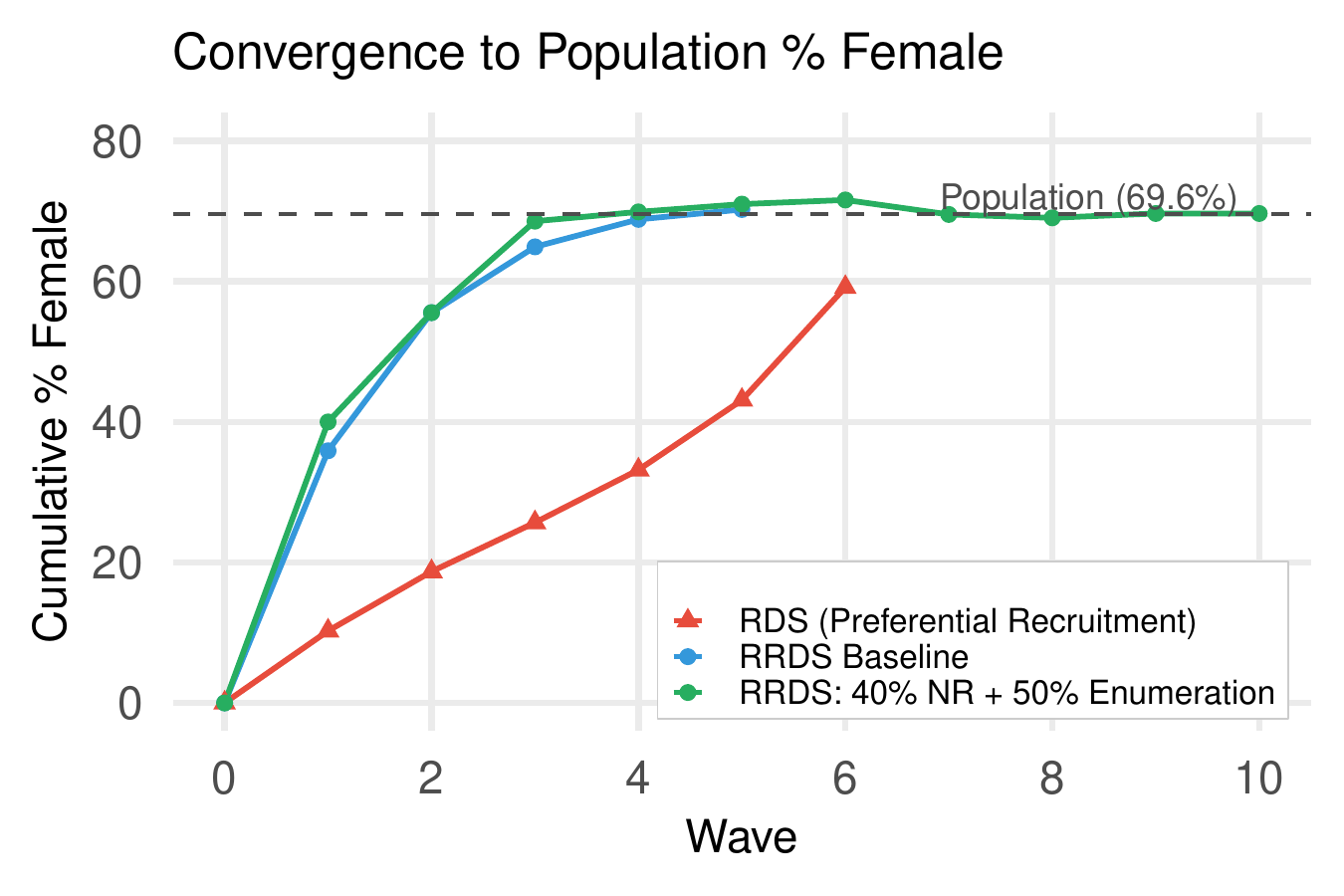}
  \caption{Simulation results comparing RDS with preferential recruitment ($\alpha_{\text{rec}} = 0.90$), RRDS, and RRDS under combined worst-case conditions (40\% non-response, 50\% contact enumeration). The network has 10{,}000 nodes, mean degree 10, and structural homophily $\alpha_{\text{net}} = 0.40$. Chains start from 10 all-male young seeds and stop at $n = 3000$. Dashed horizontal lines indicate true population values. The left figure plots cumulative mean age while the right figure plots cumulative proportion female. RRDS converges to known population parameters faster than RDS across both outcomes. Under worst-case conditions RRDS remains unbiased, though convergence is slower than the RRDS baseline.}
  \label{fig:main_sim}
\end{figure}

Figure~\ref{fig:main_sim} displays cumulative mean age and proportion female as a function of recruitment wave, for RDS, RRDS, and the combined worst-case RRDS scenario (40\% non-response and 50\% contact enumeration).

\paragraph{Bias.}
RRDS converges to population values substantially faster than RDS with preferential recruitment across both outcomes. For gender proportion -- the parameter most sensitive to preferential recruitment -- RDS remains considerably below the true population proportion (70\% female) throughout the chain, reflecting the compounding effect of $\alpha_{\text{rec}} = 0.90$ keeping chains within male subgroups. RRDS, by randomly sampling across all listed contacts, traverses gender homophily at the rate determined by the network's structure and therefore recovers more efficiently.

\paragraph{Efficiency.}
Both methods reach a cumulative sample size of $n = 3000$ within a comparable number of waves. RRDS reaches the target sample size in slightly fewer waves because random recruitment avoids the redundant nomination of previously sampled high-degree nodes that accumulates under preferential selection.

\paragraph{VH estimator.}
We examined whether the Volz-Heckathorn (VH) inverse-degree-weighted estimator \citep{volz2008probability} altered these conclusions. The VH estimator corrects for the fact that higher-degree nodes are more likely to be sampled in chain-referral designs by down-weighting such nodes proportionally to their degree. We found that the VH estimator converges slightly faster than the naive sample mean in the earliest waves, but the two are nearly indistinguishable by wave 3--4 and produce the same substantive conclusions regarding RRDS versus RDS performance. Results are presented in Appendix~\ref{app:robustness}.

\subsection{Robustness to Non-response and Partial Enumeration}
\label{sec:sim_robustness}

A practical concern for RRDS is that participants may not enumerate their full contact list and that nominated contacts may decline to participate. We examined four robustness scenarios: 20\% and 40\% non-response (nominated contacts who decline regardless of characteristics), and 70\% and 50\% partial enumeration (participants list only a random subset of their contacts). We also examined two systematic listing biases -- participants over-listing same-gender contacts, or preferentially naming higher-degree contacts -- and a combined worst-case condition of 40\% non-response with 50\% enumeration simultaneously.

The consistent finding across all scenarios is that non-response and incomplete enumeration reduce the number of effective recruits per wave, slowing the chain's traversal of the network. However, because RRDS recruits randomly from whatever contacts are listed, these limitations introduce delay but not systematic bias: all RRDS robustness scenarios eventually converge toward population truth, and all remain markedly better than RDS with preferential recruitment throughout. Reaching the target sample of $n = 3000$ simply requires more waves under challenging conditions. Full results are presented in Appendix~\ref{app:robustness}.

\section{Empirical Experiment: Garment Workers in a Pandemic}
\label{sec:data}

While garment workers in Bangladesh are not a historically hidden population, the COVID-19 pandemic made many populations ``hard-to-reach'' by removing physical access to worksites and workers' homes. This setting is particularly valuable for methodological testing; unlike most RDS applications where population parameters are unknown, we can evaluate network-based sampling performance against a known baseline derived from recent, representative surveys. This allows for a rare direct assessment of how well network-based sampling can recover true population characteristics under conditions of restricted access. We focus on the network among male respondents because we found that the network among female respondents was too highly disconnected. This sparse network structure presents a challenge to network based methods, like RRDS, as connectivity remains a core assumption. We discuss this limitation in further detail in the Discussion, Section \ref{sec:discussion}. 

The data collection process incorporated a parallel testing design, running both traditional RDS and RRDS arms simultaneously with the same starting conditions. We began with seed participants drawn from two previous traditional surveys of garment workers: a 2017 survey conducted by the Bangladesh Institute of Governance and Development (BIGD) that included 1,500 workers, and a smaller pilot survey from January 2020. The 2017 survey sample was randomly selected (by selecting every Nth household) of garment-producing areas in the Dhaka division, which encompasses approximately 80\% of Bangladesh's garment factories.

Unlike standard RDS, which only incentivizes the outcome of recruitment (a peer joining the study), we incentivize the disclosure of the recruitment network $L_i$ with a per-contact piece-rate incentive. We implement \textit{approximately-exhaustive nomination} as described in Section \ref{sec:rrds} by first asking respondents if they speak to other garment workers on the phone (89\% of men and 71\% of women do), and if so, asking respondents to provide all phone numbers from their recent call log corresponding to contacts in the industry: ``We are interested in calling workers who you have ever spoken to on the phone, that is, whose numbers are in your recent call log.'' Of respondents who reported speaking to other garment workers, 70\% of men and 63\% of women agreed to give phone numbers of their network.  Respondents received a small monetary incentive (10 taka, approximately \$0.12 US) for each phone number provided, encouraging maximal disclosure of their eligible network $N_i$. On average, men who agreed to give referrals gave 4.4 referrals and women who agreed to give referrals gave 3.3 referrals.  This approach thereby yields nomination lists $L_i$ that approximate the full set of active network ties reachable by phone.\footnote{We have a validation of the extent to which we got the full network because, before eliciting referrals, we asked workers directly how many garment workers they speak to on the phone.  On average, men gave referrals of 74\% of their network and women gave referrals of 76\% of their network.  In a subsequent study done by Boudreau and Heath that used RRDS, also in the context of the Bangladeshi garment industry, the number of referrals of women increased after the reward for referring women was increased to 50 taka, indicating that increased payments could be useful for getting these percentages even higher.  There was no corresponding decrease in the probability of success, suggesting that higher payments did not lead respondents to give made-up referrals.} Two scope conditions of this elicitation deserve note. First, phone-log ties are a subset of each respondent's full social network, so both arms sample from the phone-reachable stratum of the occupational network.  So, to the extent that this stratum is itself more homogeneous than the full network, the observed differences between the RDS and RRDS arms understate the differences that would arise over complete networks. Second, the call log covers a finite recall window, so recently inactive ties are excluded from $L_i$ in both arms.

Sixty three percent of referrals selected to be surveyed were successfully surveyed; success rates were higher among referred women (72\% of referred women were successfully surveyed, versus 58\% of referred men) and among the RDS ``preferred'' sample. (72\% of ``preferred'' seeds were successfully surveyed, versus 62\% of seeds in the RRDS arm).

The two study arms differ only in the recruitment stage. In the traditional RDS arm, respondents were asked which contacts they felt most comfortable referring, and these preferred contacts were recruited -- allowing respondent-driven selection.\footnote{That is, to avoid introducing recruitment bias, we did not provide incentives based on successful survey completion. Our ultimate finding that RRDS performs better than RDS is thus likely an underestimate of the true difference between the groups, given that the RDS estimate in our case reflects homophily in the preferred referrals but not any additional selectivity based on the selection of network members most likely to complete the survey.} In the RRDS arm, contacts were randomly selected from the provided list without regard to the respondent's stated preferences, ensuring uniform sampling from $L_i$. All selected participants were given a brief ``intermediate link'' survey eliciting demographics, health, and recent workplace outcomes, allowing us to track the characteristics of workers recruited under RDS and RRDS, then were asked to give referrals of their own using the same procedure described above.  This experimental design provides empirical validation of the simulation findings while also revealing practical considerations for implementation. 

\section{Results}
\label{sec:results}

We present results for male respondents only, as low female recruitment prevented valid inference. This outcome reflects the disconnected network structure of the target population rather than a limitation of the randomized recruitment protocol (see Section \ref{sec:discussion}). We note that restricting the analytic sample by gender adds an assumption beyond the standard RDS framework: our estimands are properties of the male garment-worker population, inference proceeds on the male sub-network, and cross-gender ties followed during data collection are treated as censored. Standard RDS assumptions are not guaranteed to hold on a subsample defined post hoc, and estimates should be interpreted accordingly. 

We estimate mean age and proportions for key demographic characteristics (Dhaka-born, married, primary school completion, secondary school completion, and living with a child under age 5) from both RDS and RRDS samples. For each recruitment strategy, we calculate estimates using two estimators: traditional Volz-Heckathorn RDS weights \citep{volz2008probability}, described in Section \ref{sec:rds}, with confidence intervals constructed via the standard RDS bootstrap of \citet{salganik2006variance} (labeled ``Traditional'' in the figures), and the RDS Tree Bootstrap \citep{baraff2016estimating}, whose point estimate is the VH estimate computed on the original sample (Algorithm~\ref{alg:treeboot}, line~2) with intervals taken from the weighted percentiles of the bootstrap distribution. The Tree Bootstrap builds upon the VH estimator in Equation \ref{eq: VH}, using the process described in Algorithm \ref{alg:treeboot} which estimates uncertainty by simulating the recruitment process itself rather than treating respondents as independent data points. It works by resampling hierarchically: first, it randomly selects initial ``seeds" with replacement, and then for each selected person, it randomly draws from their actual set of recruits to continue the chain. This preserves the dependency structure of the data—ensuring that a respondent is only included if their recruiter was included—which accurately reflects the clustering and ``chain-like" nature of RDS. While we present both for comparison, we recommend the Tree Bootstrap weights as they better capture the uncertainty inherent in network-based sampling. We validate these 2020 estimates against the \citet{kabeer2019multi} survey of 1,500 garment workers collected in 2017 to assess recruitment effectiveness. This is an imperfect benchmark (i.e. things may have changed since 2017), but it is useful to have a recent point of reference. Such validation against estimates from traditional survey sampling is not feasible in studies of sensitive or hard-to-reach populations using RDS, so this comparison offers one of the few opportunities to gauge the plausibility and direction of network-based survey estimates.

\begin{algorithm}
\caption{Tree Bootstrap for RDS Uncertainty Estimation \citep{baraff2016estimating}}
\label{alg:treeboot}
\begin{algorithmic}[1]
\REQUIRE Observed RDS trees $\mathcal{T} = \{T_1, \ldots, T_S\}$ with sample $\mathcal{S}$
\REQUIRE Number of bootstrap samples $B$
\ENSURE Bootstrap distribution of VH estimator $\{\hat{\mu}_{\text{VH}}(\mathcal{S}^*_1), \ldots, \hat{\mu}_{\text{VH}}(\mathcal{S}^*_B)\}$

\STATE \textbf{Compute point estimate from original sample:}
\STATE $\hat{\mu}_{\text{VH}}(\mathcal{S}) = \dfrac{\sum_{i \in \mathcal{S}} d_i^{-1} x_i}{\sum_{i \in \mathcal{S}} d_i^{-1}}$ \COMMENT{VH estimator on original sample}
\vspace{0.2cm}
\FOR{$b = 1, \ldots, B$}
    \STATE $\mathcal{S}^*_b \gets \emptyset$ \COMMENT{Initialize bootstrap sample}
    
    \STATE \textbf{Resample seeds:}
    \STATE Sample $S$ seeds with replacement from $\{T_1, \ldots, T_S\}$ $\rightarrow$ add to $\mathcal{S}^*_b$
    
    \STATE \textbf{Recursively resample descendants:}
    \FOR{each level $\ell = 0, 1, 2, \ldots$ until no children remain}
        \FOR{each node $s$ at level $\ell$ in bootstrap sample}
            \STATE $C(s) \gets$ children of $s$ in original tree
            \IF{$|C(s)| > 0$}
                \STATE Sample $|C(s)|$ children with replacement from $C(s)$
                \STATE Add to $\mathcal{S}^*_b$ as descendants of $s$
            \ENDIF
        \ENDFOR
    \ENDFOR
    
    \STATE \textbf{Apply VH estimator to bootstrap sample:}
    \STATE $\hat{\mu}_{\text{VH}}(\mathcal{S}^*_b) = \dfrac{\sum_{i \in \mathcal{S}^*_b} d_i^{-1} x_i}{\sum_{i \in \mathcal{S}^*_b} d_i^{-1}}$ \COMMENT{Same formula, different sample}
\ENDFOR

\STATE \textbf{Construct confidence interval:}
\STATE Let $w_b = \sum_{i \in \mathcal{S}^*_b} d_i^{-1}$ be the effective sample size for bootstrap sample $b$
\STATE Compute weighted $(1-\alpha)$ CI using percentiles of $\{\hat{\mu}_{\text{VH}}(\mathcal{S}^*_b)\}_{b=1}^B$ with weights $\{w_b\}_{b=1}^B$
\end{algorithmic}
\end{algorithm}

To assess overall performance across variables, we calculate standardized Root Mean Square Error (RMSE) to measure bias relative to the 2017 baseline survey. To evaluate uncertainty quantification, we calculate the coverage rate -- the proportion of 95\% confidence intervals that contain the true baseline estimate for each demographic variable. The RRDS recruitment strategy with the Tree Bootstrap estimator produces the least biased estimates compared to 2017 (RMSE = 2.503) and has the best coverage of the underlying population (83.3\% of confidence intervals contain the baseline estimate). A full table of comparisons are shown in Table \ref{tab:recruitment_estimator_comparison}. Traditional RDS performs particularly poorly, with zero coverage when using traditional Volz-Heckathorn weights, indicating that confidence intervals consistently fail to capture the 2017 assumed-to-be-true population parameters. The Tree Bootstrap estimator improves performance for both recruitment methods, but the combination of RRDS recruitment with Tree Bootstrap weighting provides the most substantial gains in both bias reduction and uncertainty quantification. 

\begin{figure}[!ht]
    \centering
    \includegraphics[width=0.48\textwidth]{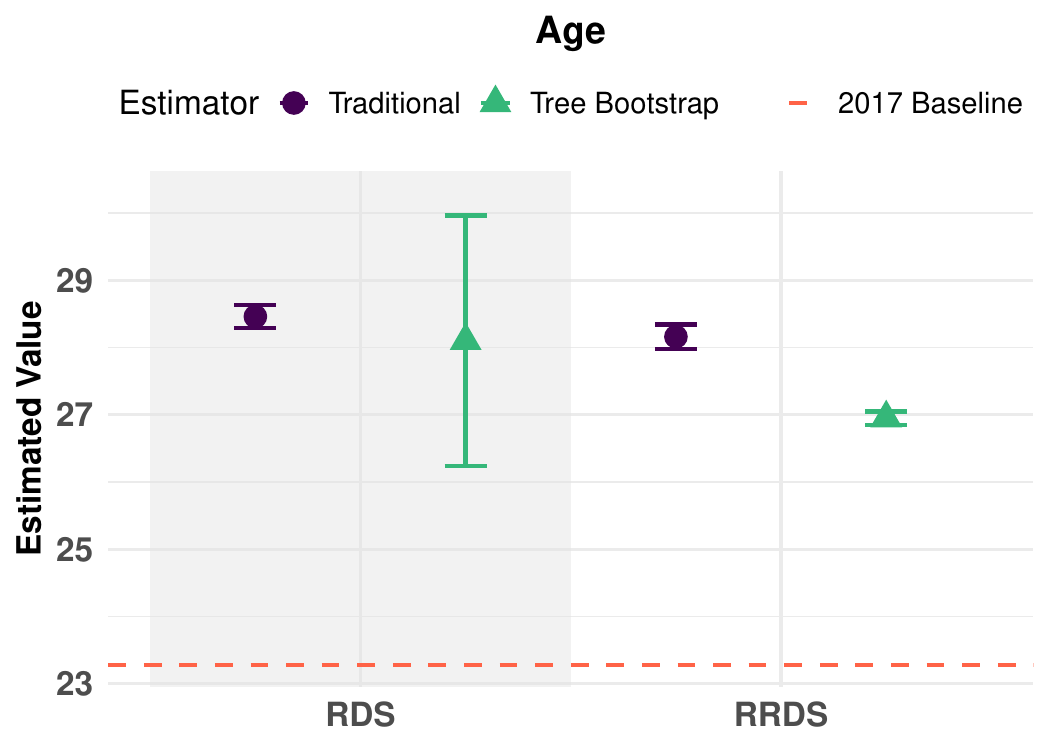}
    \includegraphics[width=0.48\textwidth]{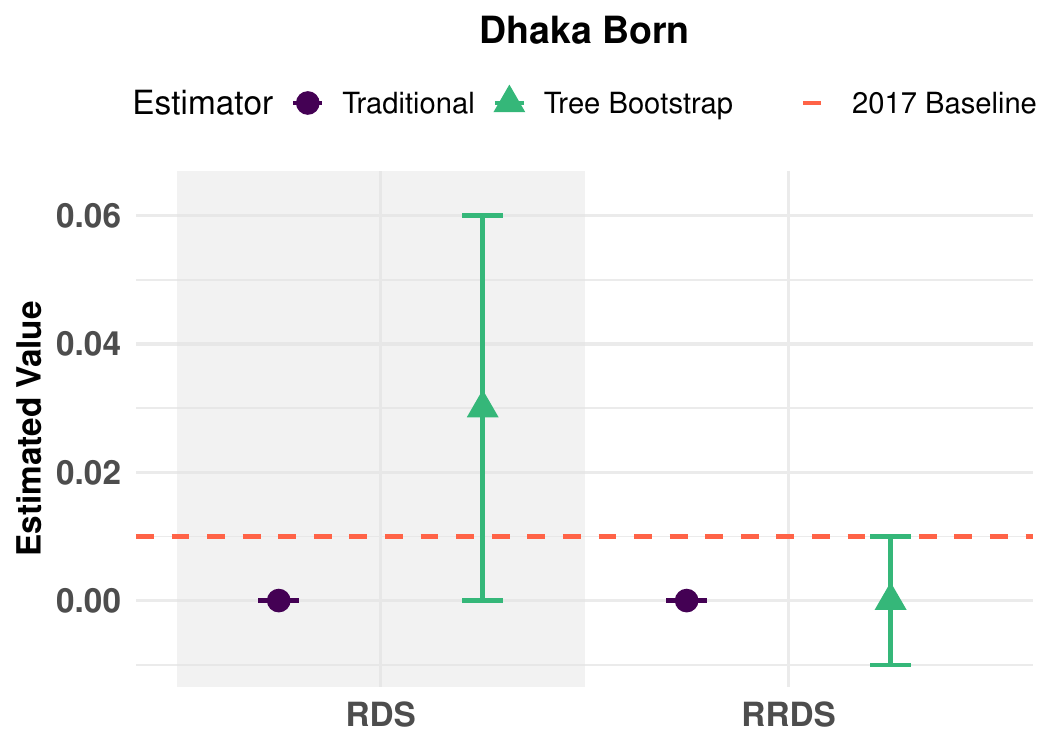} \\
    \includegraphics[width=0.48\textwidth]{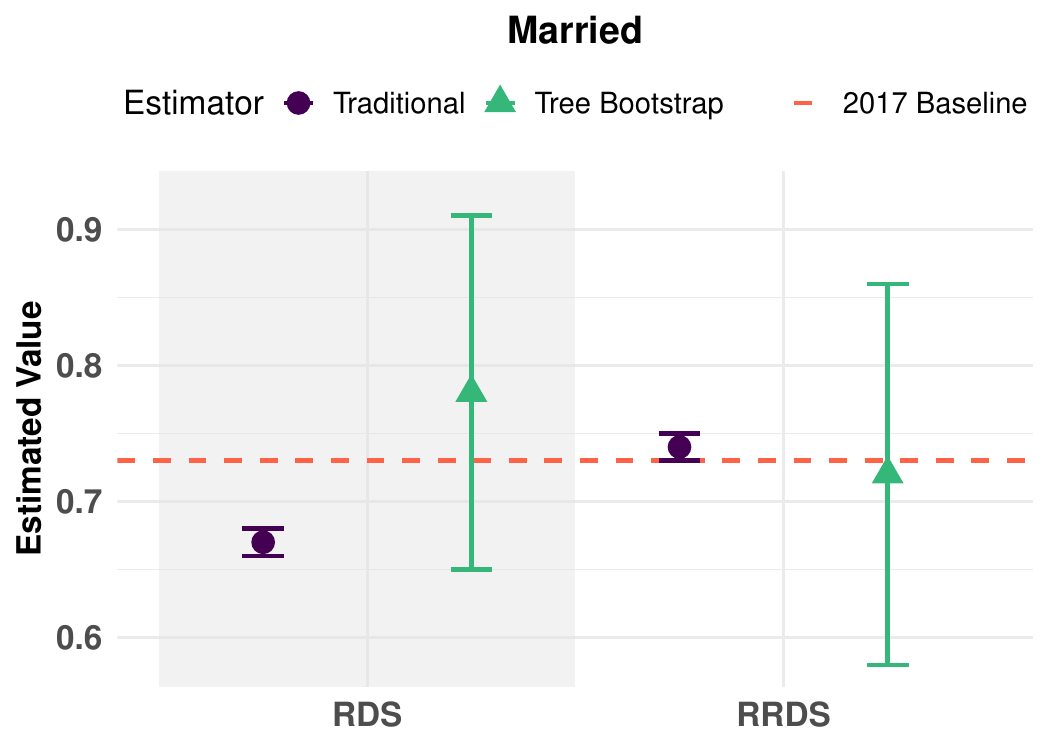}
    \includegraphics[width=0.48\textwidth]{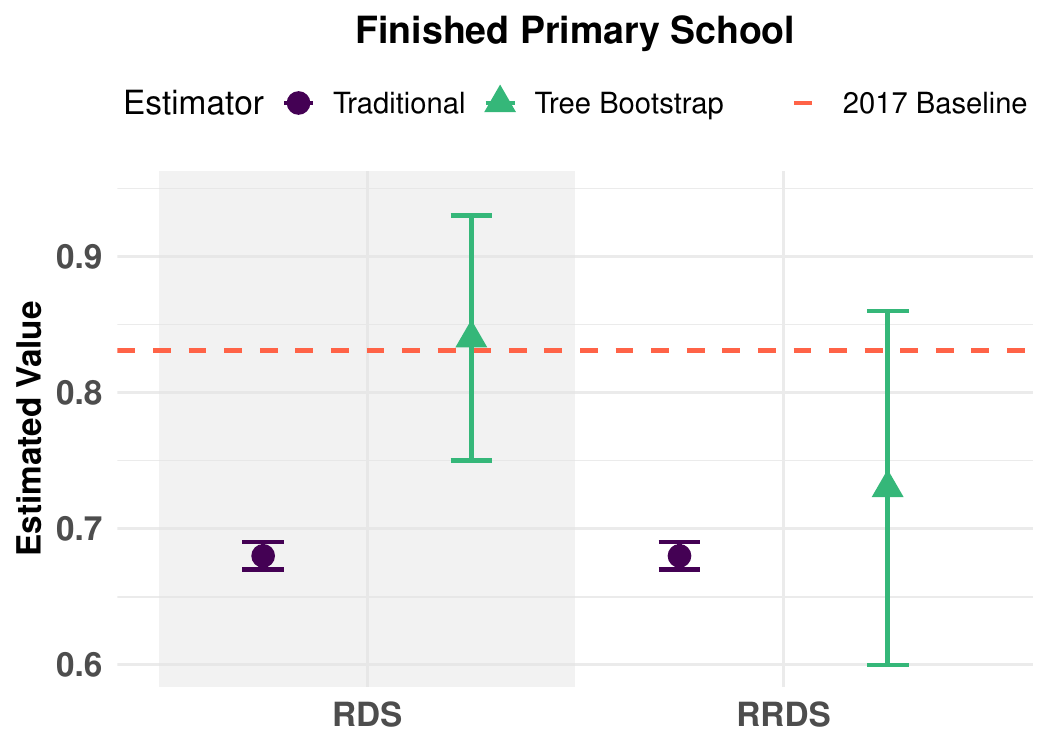} \\
    \includegraphics[width=0.48\textwidth]{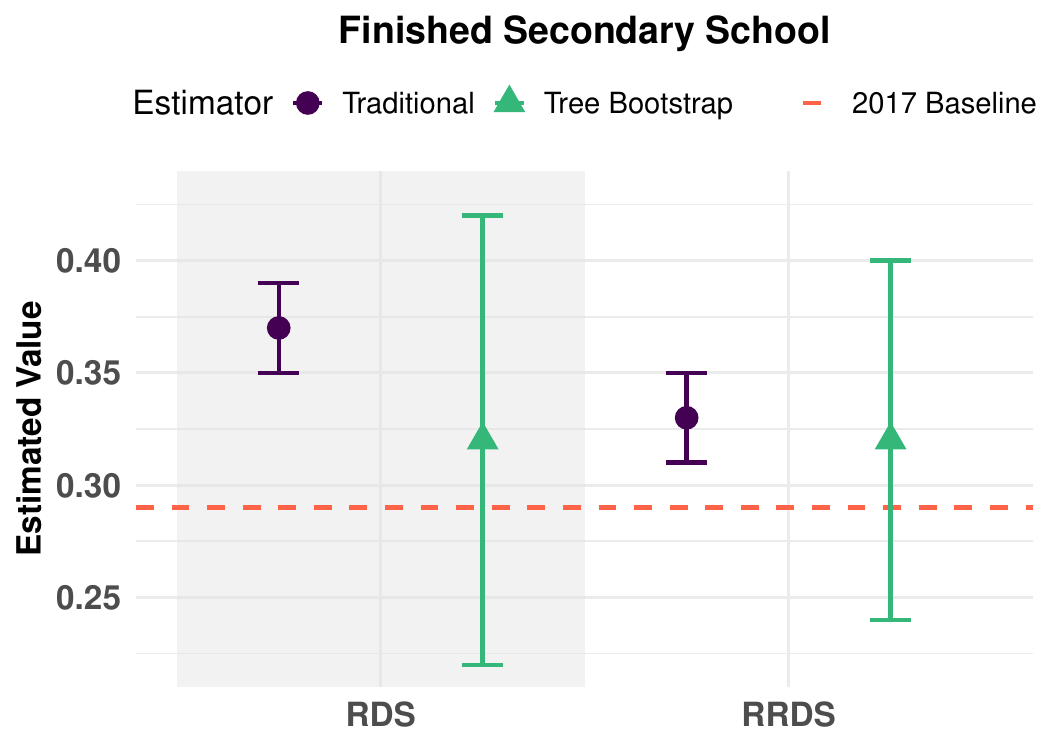}
    \includegraphics[width=0.48\textwidth]{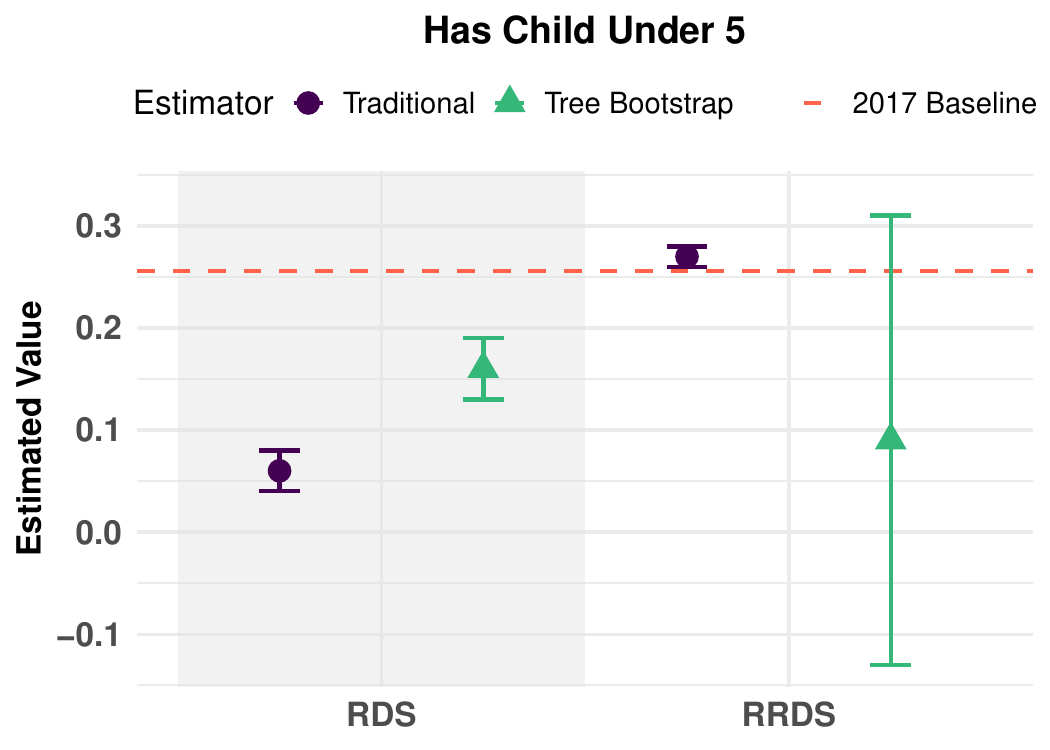}
    \caption{Comparison of RDS and RRDS estimates across six demographic characteristics. The Tree Bootstrap estimator generally reflects greater uncertainty due to network structure and sampling variability. Randomized recruitment tends to yield less biased estimates.}
    \label{fig:demographic_grid}
\end{figure}

These results indicate that combining RRDS recruitment with Tree Bootstrap weighting substantially improves both point estimation accuracy and the validity of uncertainty intervals relative to the 2017 baseline. When aggregating performance across the six demographic parameters, the RRDS estimator with Tree Bootstrap weights achieves the highest observed coverage: its 95\% confidence intervals capture the baseline population parameter for five of the six variables (83.3\%). While this represents a marked improvement over traditional RDS methods, the fact that the interval still misses the baseline for one variable suggests that fully accounting for the complex design effects of network-based sampling remains a challenge. Variable-specific comparisons are visualized in Figure \ref{fig:demographic_grid} and presented in Table \ref{tab:recruitment_estimator_comparison}.

\begin{table}[ht]
\centering
\caption{Comparison of Recruitment Methods and Estimators. The RRDS recruitment strategy with the Tree Bootstrap estimator produces the least biased estimates and has the best coverage (95\% CI containing the baseline) of the underlying population.}
\vspace{0.4cm}
\label{tab:recruitment_estimator_comparison}
\begin{tabular}{@{}llcc@{}}
\toprule
\textbf{Recruitment} & \textbf{Estimator} & \textbf{Bias (RMSE)} & \textbf{CI Coverage (Count/6)} \\
\midrule
RDS  & Volz-Heckathorn & 3.713 & 0/6 \\
RDS  & Tree Bootstrap  & 3.118 & 4/6 \\
RRDS & Volz-Heckathorn & 3.198 & 1/6 \\
RRDS & Tree Bootstrap  & \textbf{2.503} & \textbf{5/6} \\
\bottomrule
\end{tabular}
\end{table}

For marriage status, RRDS yields substantially less biased estimates than traditional RDS. When traditional RDS weights are applied, the RRDS estimate is not statistically different from the true population parameter, while the RDS estimate shows significant deviation. This suggests that the randomization step effectively mitigates homophily-based selection bias for this characteristic.

Educational attainment highlights differences between weighting strategies. For primary school completion, the two recruitment methods are essentially indistinguishable under traditional weights: both point estimates fall significantly below the 2017 baseline. Under Tree Bootstrap weights, the RDS point estimate happens to land closer to the baseline than the RRDS estimate, although both intervals cover the baseline value; primary schooling is thus the one characteristic for which RRDS does not outperform RDS. For secondary school completion, RRDS shows less bias than RDS under both weighting methods, though both are positively biased relative to the baseline. Across both education variables, Tree Bootstrap weights produce substantially wider confidence intervals than traditional weights, consistent with the method's purpose of capturing dependence-induced uncertainty.

The proportion of respondents with children under 5 years old reveals particularly strong homophily effects in traditional RDS. While the RRDS estimate using Tree Bootstrap weights shows considerable uncertainty, the RRDS estimate with traditional weights is far less biased than the RDS estimate. This demonstrates RRDS's effectiveness in counteracting homophily-driven recruitment patterns in family structure characteristics.

For the Dhaka-born proportion, both methods struggle to produce reliable estimates due to the small baseline proportion in the population and limited sample size of the network-based sampling. This highlights an important limitation: very rare population characteristics may remain difficult to estimate precisely even with methodological improvements to RDS.

Taken together, RRDS performs comparably to or better than traditional RDS on every demographic dimension examined except primary school completion, where RDS with Tree Bootstrap weights yields the point estimate closest to the 2017 baseline. The consistency of these improvements across different demographic characteristics suggests that the randomization mechanism effectively addresses non-random selection bias while maintaining the practical advantages of network-based recruitment. These empirical results complement and reinforce the findings from the simulation in Section \ref{sec:sim}, providing real-world validation of RRDS's theoretical advantages.

\section{Discussion}
\label{sec:discussion}

Non-probability sampling is increasingly common in social science research because obtaining probability samples has become prohibitively difficult and expensive \citep{Breen2025_newdata}. This methodological shift is particularly relevant to disaster settings, like during COVID-19, where the logistical hurdles that hinder standard sampling are amplified by chaos and displacement, making non-probability designs an appealing alternative. Randomized Recruitment Driven Sampling (RRDS) represents a methodological bridge between conventional probability sampling and network-based recruitment, designed for settings where logistical constraints—rather than population stigma—preclude standard survey methods. Unlike traditional RDS, which was developed for stigmatized populations where full network elicitation is infeasible, RRDS leverages the full network and introduces researcher-controlled randomization into the recruitment process \citep{rds_real_world, assess_rds}.

The simulation study suggests that RRDS produces more representative samples, converging to population parameters in fewer waves than traditional RDS with preferential recruitment. However, the empirical application revealed both the strengths and the inherent structural limits of network-based inference. While RRDS is more efficient than traditional RDS at exploring a connected component, it remains, fundamentally, a network-based method. If the underlying social graph is sparse or disconnected -- as observed here among female workers in particular -- recruitment chains will inevitably break. The limitation in this context, therefore, lies not in the randomization mechanism of RRDS, but in the lack of sufficient network connectivity required to sustain any chain-referral approach.
Several factors likely contributed to women's reduced willingness or ability to provide referrals. 

First, women are limited in their ability to maintain independent social networks that can be mobilized for survey recruitment.  While male garment workers frequently socialize with colleagues outside work hours, female workers often return directly home after shifts due to safety concerns and family responsibilities.  Moreover, women's cell phone access is constrained, given they often have to borrow a phone from a husband or male family member or face oversight of a phone that is nominally their own \citep{boudreau2024migrants}. Both factors constrain women's opportunities to form and maintain the broad occupational networks that facilitate successful referral chains in network-based recruitment methodologies.

Second, the 2013 Rana Plaza factory collapse, which killed over 1,100 garment workers in Bangladesh, intensified global scrutiny of labor practices and worker safety in the garment industry. Female workers, who typically occupy more vulnerable positions in factory hierarchies, may hesitate to refer colleagues into a survey addressing potentially sensitive topics out of fear of workplace repercussions \citep{boudreau2023monitoring}. As one female respondent noted during recruitment, ``I know other workers, but I don't want to create problems for them.''

Future research should consider weighted sampling techniques (i.e., choosing networks referrals most unlike the referrer with higher probability) to achieve convergence faster, examine the performance of RRDS across different network topologies, and develop specialized variance estimators that account for the unique features of randomized network recruitment. Our field adaptation of exhaustively following all female contacts represents an extreme instance of this concept, underscoring the value of formalizing such adaptive strategies through frameworks like Network Sampling with Memory \citep{Mouw2012NSM}. Additionally, expanding the approach to incorporate digital recruitment channels may further enhance its applicability in increasingly connected populations, while remaining attentive to gender-based digital divides.

In summary, RRDS represents a valuable addition to the methodological toolkit for researchers working with populations that are difficult to sample due to logistical rather than social barriers. It is not necessarily a replacement for traditional RDS in studies of stigmatized or hidden populations, where participants might be hesitant or unable to provide the full network elicitation that RRDS requires. Rather, RRDS fills a specific niche: settings such as disaster zones, conflict areas, or pandemic conditions where populations are temporarily inaccessible but not subject to stigma that would prevent network disclosure. In these contexts, traditional RDS could also be employed, but RRDS offers the advantage of satisfying the random recruitment assumption that RDS requires but rarely achieves in practice \citep{heckathorn2017, raifman_2022}, offering a pragmatic compromise between statistical rigor and practical feasibility in challenging research environments. However, its implementation must be attentive to contextual factors, particularly gender dynamics, that shape referral patterns and ultimately determine sample composition.

\section{Acknowledgments}
AV and TM were supported by the National Institute of Mental Health of the NIH under Award Number DP2MH122405 and by the Center for Statistics and the Social Sciences at the University of Washington. AV gratefully acknowledges the resources provided by the International Max Planck Research School for Population, Health and Data Science (IMPRS-PHDS). The content is solely the responsibility of the authors and does not necessarily represent the official views of the NIH.

\newpage
\bibliography{cites}

\newpage

\appendix

\section{Appendix}

\subsection{Simulation Robustness Checks}
\label{app:robustness}

\begin{figure}[!ht]
  \centering
  \includegraphics[width=\textwidth]{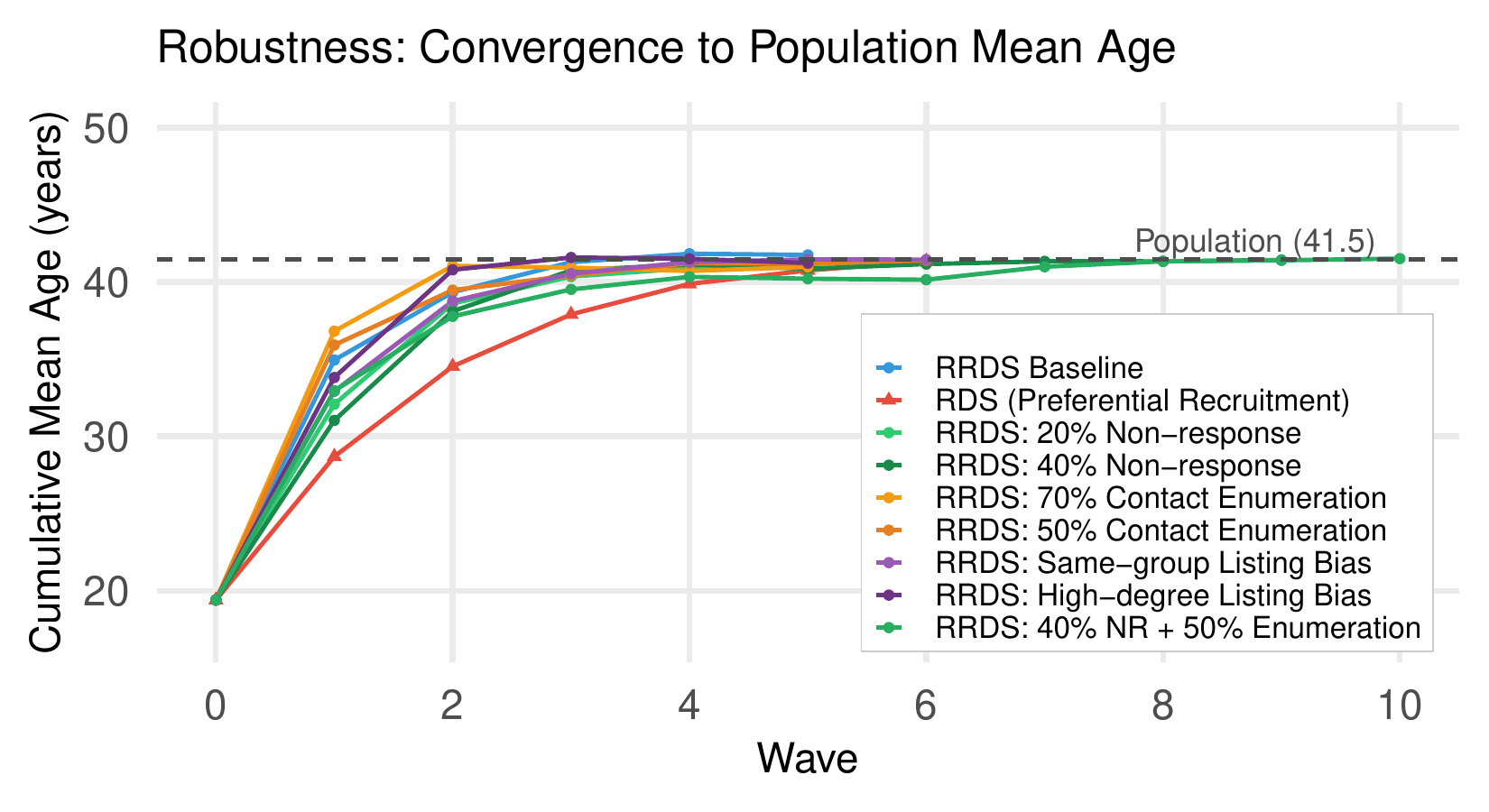}
  \caption{Robustness simulation: cumulative mean age by wave across all scenarios. Even under the most challenging conditions -- 40\% non-response combined with 50\% contact enumeration -- RRDS converges toward the true population mean age more reliably than RDS with preferential recruitment. Non-response and partial enumeration slow convergence (chains require more waves to reach the target sample size) but do not introduce systematic bias.}
  \label{fig:rob_age}
\end{figure}

\begin{figure}[!ht]
  \centering
  \includegraphics[width=\textwidth]{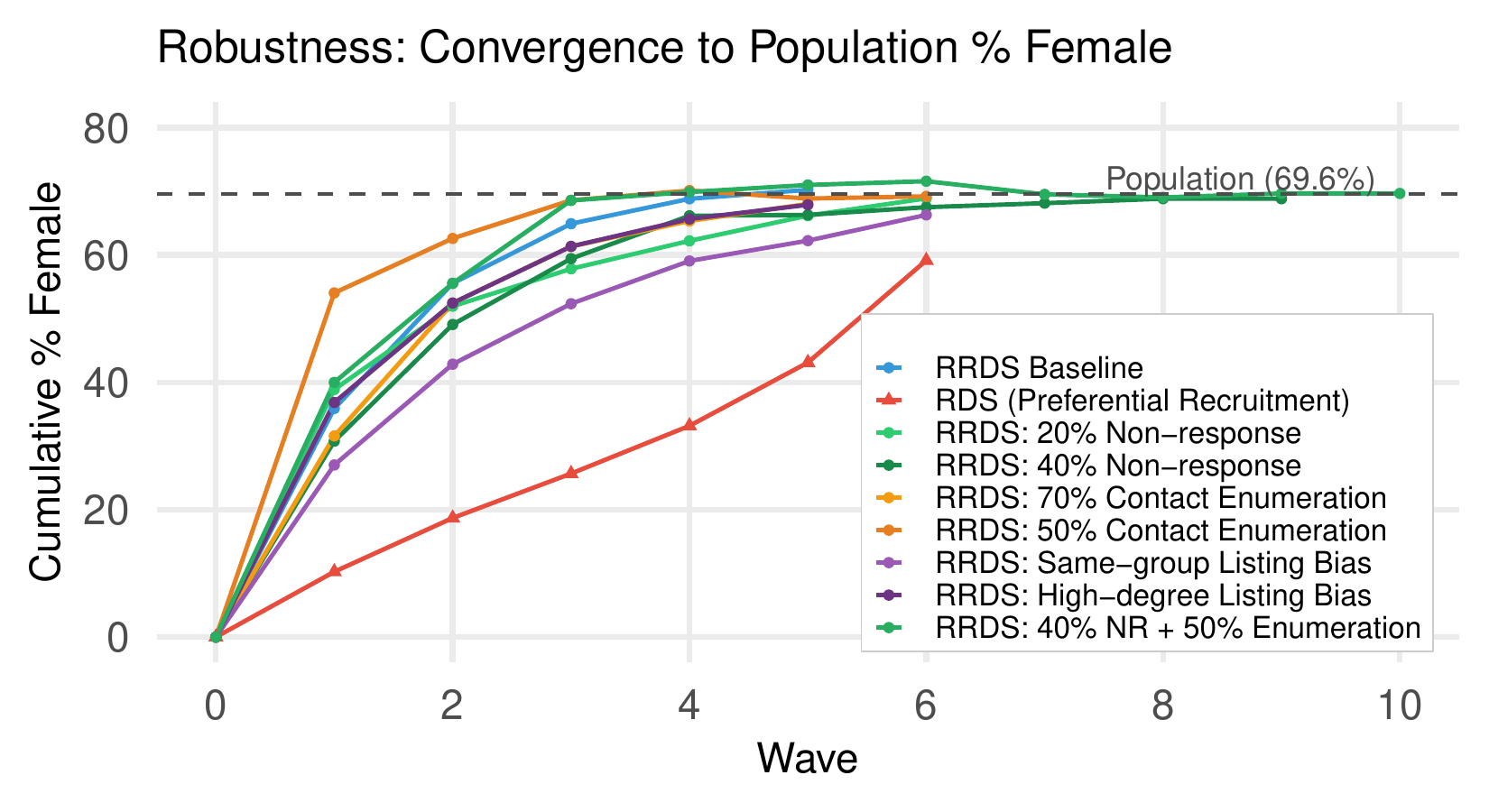}
  \caption{Robustness simulation: cumulative proportion female by wave across all scenarios. The gender proportion result is the most sensitive to preferential recruitment bias and therefore the clearest demonstration of RRDS's advantage. Under all robustness conditions, RRDS tracks toward the true population proportion (70\% female) considerably faster than RDS. The same-group listing bias scenario -- where participants disproportionately enumerate same-gender contacts -- produces the greatest slowing among RRDS variants, yet still outperforms RDS.}
  \label{fig:rob_female}
\end{figure}

Figures~\ref{fig:rob_age} and~\ref{fig:rob_female} display cumulative mean age and proportion female for all simulation scenarios. In addition to the baseline RDS and RRDS, results are shown for RRDS under 20\% and 40\% non-response, 70\% and 50\% partial contact enumeration, same-group and high-degree listing biases, and the combined 40\% non-response plus 50\% enumeration worst case.

Here we share further detail about the several waves of data collection. 

\subsection{Wave 1 - Seed Sample} 

Seeds for referral chains were collected from two prior surveys:

\begin{enumerate}
    \item Administered in 2017, 1500 workers were surveyed by the Bangladesh Institute of Governance and Development (BIGD) and London School of Economics \citep{kabeer2019multi}.
    \item Administered in 2020, 60 workers were surveyed in a pilot survey.
\end{enumerate}

These two samples include both the names and phone numbers of respondents and, when originally recruited, were geographically representative of garment-producing areas in the Dhaka Division of Bangladesh, home to 80\% of the country’s garment factories. 

The data from the 2017 BIGD survey were stratified by cells defined by location (evenly split between Ashulia, Gazipur, Narayanganj, Mirpur, and Hemayetpur), gender, and above or below median experience (for one’s given gender-location group).

\subsection{Intermediate Links}
All wave 1 seed participants were asked to recruit their contacts using the following script: 

\begin{quote}
    \emph{We would like to survey more garment workers to learn about their experiences. We are interested in calling workers who you have ever spoken to on the phone, that is, whose numbers are in your recent call log. We will provide you BDT 10 for each worker (i.e., a name and phone number) that you give us, for up to 10 workers total. We will then use a lottery to choose one or more of them to call. They may be selected to answer a much shorter survey than the one you just received, which will take about 10 minutes, and receive BDT 50 as a thank - you for their time. We would let them know that you provided a referral. Would you be willing to provide the names and phone numbers of family members and/or friends who work in the garments sector who may be willing to participate?} 
\end{quote}

Additionally, all wave 1 respondents - and subsequent referrals given by these respondents - were asked the following questions: 

\begin{itemize}
    \item If the respondent listed 2 or more names: 
    \begin{quote}
        \emph{Out of the workers you just listed, whom would you feel most comfortable referring?}
    \end{quote}
    
    \item If the respondent listed 3 or more names: 
    \begin{quote}
        \emph{Whom would you feel next most comfortable referring?}
    \end{quote}

    \item If the respondent listed 4 or more names: 
    \begin{quote}
        \emph{Whom would you feel next most comfortable referring?}
    \end{quote}
\end{itemize}

Referrals from these wave 1 seed respondents were randomly assigned to either “randomized” (RRDS) or “traditional” (RDS) recruitment groups according to the following protocols. 

\begin{enumerate}
    \item The first round of the intermediate links survey began on November 26, 2020. In the RRDS group, 3 contacts (or however many contacts were given, if less than 3) were picked at random to be followed and asked for referrals of their own. In the RDS group, three preferred contacts (or however many contacts were given, if less than 3) were followed.  
    \item On December 12, 2020, we began following all referrals of women, regardless of treatment group. 
    \item From December 21, 2020 until we finished the intermediate links survey on January 16, 2021,  we began following up to four referrals per respondent, using the following algorithm:
    
    a. From respondents in the RRDS group, we selected four referrals randomly.
    
    b. For respondents in the RDS group, we first selected one referral randomly from all referrals that they gave (both the preferred and others). Then, if the randomly selected referral was one of the three preferred referrals, we
    just follow the 3 preferred referrals, as we have been doing already.  If the randomly selected referral was not one of the three preferred referrals, we follow that random referral and the 3 preferred referrals, for up to four referrals total.  
\end{enumerate}

These updates to the sampling procedure were made to adjust for the behavior of respondents. Early on we observed that women were exceedingly unlikely to provide referrals despite the incentive, resulting in the expiration of those recruitment chains. As discussed in Section~\ref{sec:discussion}, this was an unforeseen challenge that we did our best to overcome. 

In total, there were 30 waves of the intermediate links survey, where the selected referrals from wave $n$ constitute the sample for referrals in wave $n+1$.

\subsection{Wave 2a}
Wave 2a consisted of a random selection of 25\% of the 2017 sample, as well as selections from intermediate links, selected in the following way:

\begin{enumerate}
    \item 25\% of factories were selected, and 3 workers per factory were selected.
    \item Up to 10 workers from wave 10 on-wards from each factory that agreed to be surveyed in the Bangladesh Garment Manufacturers and Exporters Association (BGMEA) matched survey conducted by Laura Boudreau beginning in January 2021. If a BGMEA factory appears after wave 10 but has fewer than 3 post wave 10, allow up to 3 workers using respondents from waves before 10. 
    \item All respondents not currently employed in garment factories were selected.
    \item All women from wave 10 on-wards were selected , up to a max of 15 per factory.
    \item 50\% of workers with 3 or fewer years of experience from random seeds in wave 10 on-wards were selected. 
    \item For those who have left the garment sector, select up to 3 workers from that same factory if they don’t already have at least 3 workers sampled.
    \item There was a cap of 5 men per factory (that wasn’t matched to BGMEA). So randomly throw out men if more than this number was selected.
\end{enumerate}

\subsection{Wave 2b}
Among referrals given by the respondents to wave 2a: 

\begin{enumerate}
    \item All referrals not currently employed in garment factories were selected. 
    \item All referrals of women were selected, unless they lead to greater than 15 workers per factory in expectation (assuming a 50\% probability that a selected respondent leads to a completed survey, based roughly on the intermediate links surveys).
    \item All referrals of newly agreed upon factories in the BGMEA survey. 
    \item Add men until 3 per factory in expectation are selected. Choose from among men with 3 or fewer years experience if available. 
    \item 20 men with more than 3 years experience were randomly selected.

\end{enumerate}

\subsection{Wave 2c}
Among referrals given by the respondents to wave 2b:

\begin{enumerate}
    \item Select referrals in factories in the BGMEA survey who currently have less than 10 successful surveys. Select women to yield 10 successful surveys in expectation, again assuming at 50\% success rate. 
    \item Select all women unless their factory has more than 20 current surveys. 
    \item Select 100 men randomly.

\end{enumerate}

Additional selections were made from the intermediate links:

\begin{enumerate}
    \item In factories in the BIGD survey that did not yield 10 successful surveys so far, select workers to yield 10 successful surveys in expectation. 
    \item Select women whose factories have 5 or fewer surveys currently.

\end{enumerate}

\subsection{Wave 2d}
Among referrals given by the respondents to wave 2c: 

\begin{enumerate}
    \item Select all referrals from those in BGMEA matched factories.
    \item Choose 60 new factories at random from factories not matched to BGMEA survey. Choose all women from these factories and 25\% of men in these factories.
    \item Select all women who have left the sector. 
    \item Select 25\% of men who have left the sector.

\end{enumerate}

\newpage
\section{Figures}

\begin{figure}[!ht]
    \centering
    \includegraphics[width=0.95\textwidth]{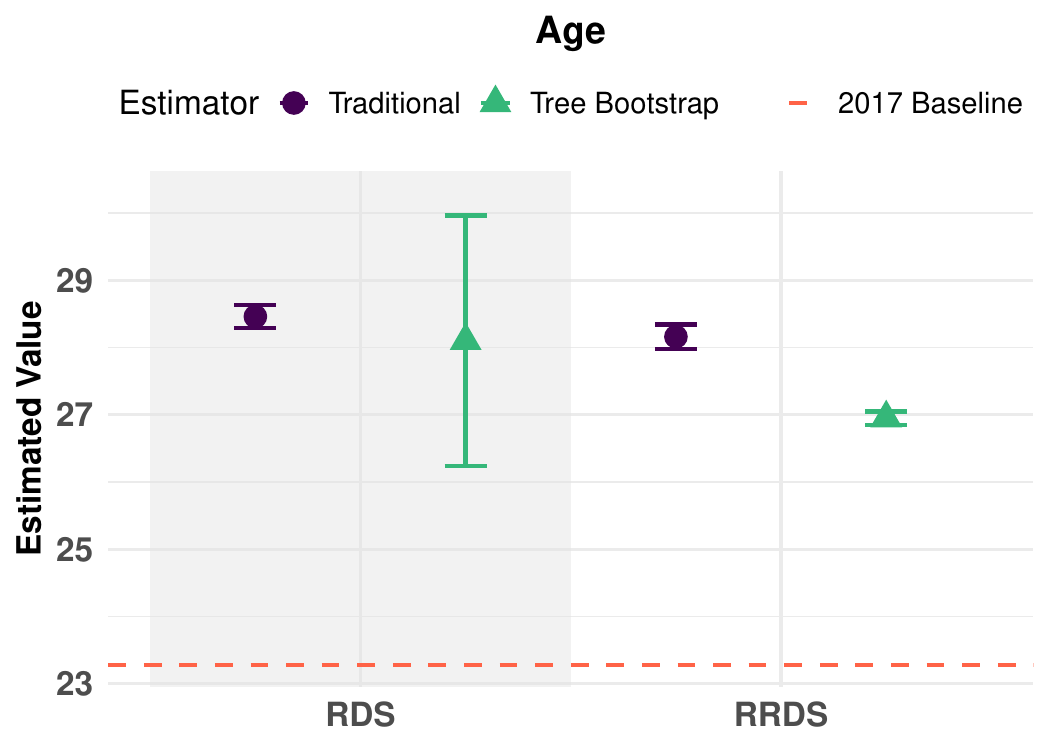}  
    \caption{Each of the RDS estimates more closely approximate the population mean age from the 2017 baseline survey. The Tree Bootstrap estimator captures additional uncertainty from sampling and network structure, leading to wider intervals than the traditional RDS estimator.}
\end{figure}

\begin{figure}[!ht]
    \centering
    \includegraphics[width=0.95\textwidth]{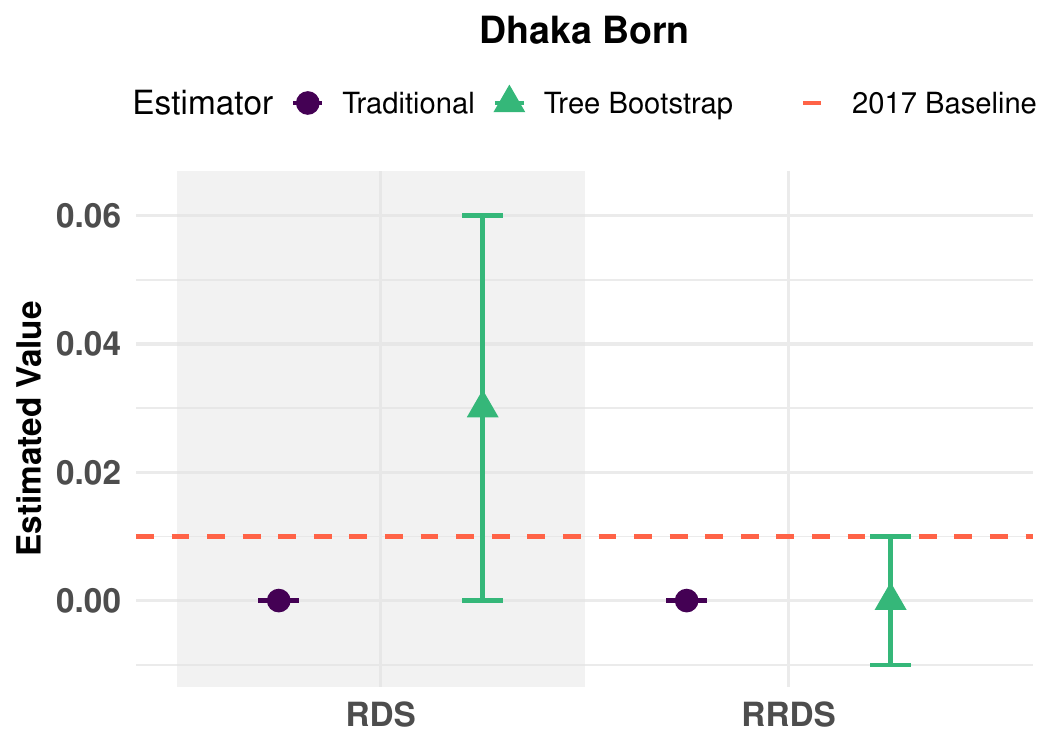}  
    \caption{Proportion of respondents who are native born to the Dhaka region across recruitment waves. The horizontal dashed line represents the population parameter from the representative 2017 survey. This is a very small proportion of the population, and due to the limited sample size from the network-based sampling it is not possible to recover a reliable estimate of this quantity with either method.}
\end{figure}

\begin{figure}[!ht]
    \centering
    \includegraphics[width=0.95\textwidth]{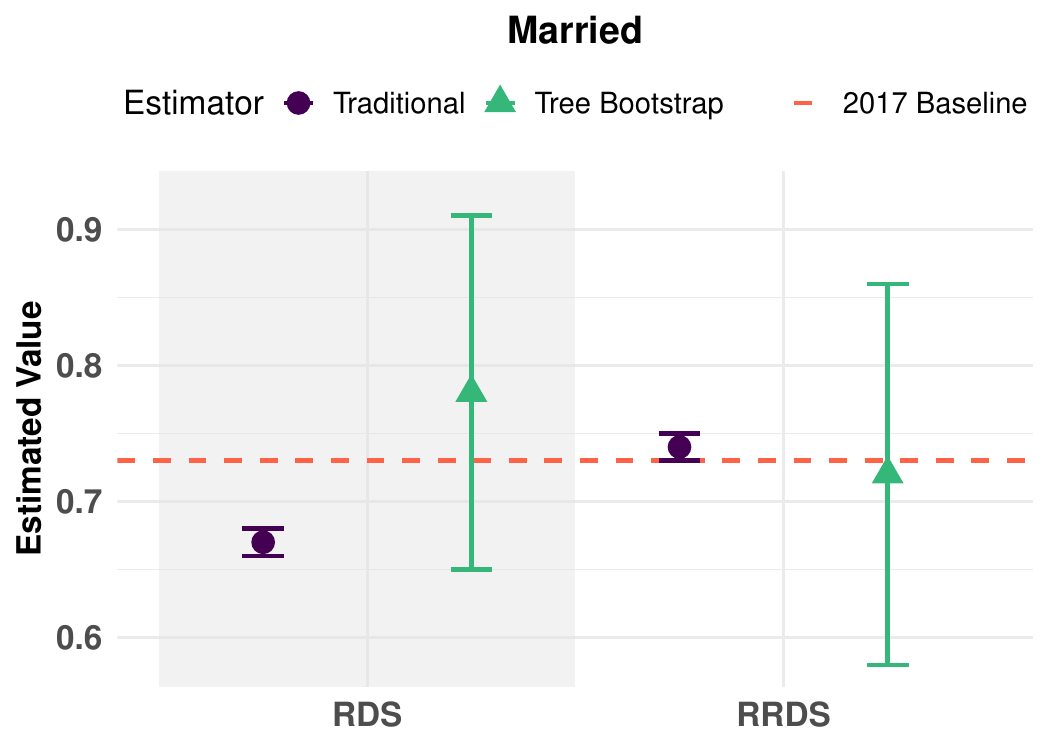}
    \caption{Proportion of married respondents across recruitment waves. RRDS point estimates are far less biased than traditional RDS. With traditional RDS weights, the RRDS estimate is not statistically different from the true population parameter, while the RDS estimate is. }
\end{figure}

\begin{figure}[!ht]
    \centering
    \includegraphics[width=0.95\textwidth]{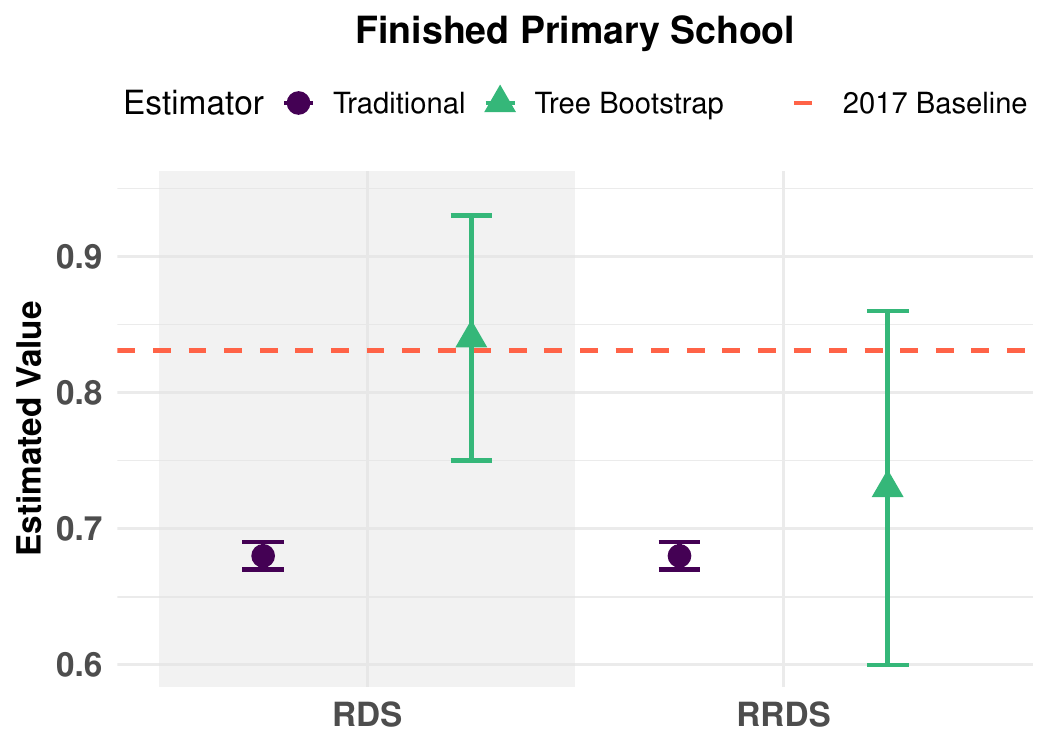}
    \caption{Proportion of respondents with primary school education across recruitment waves. The RRDS estimate with RDS Tree Bootstrap weights is more biased than the RDS estimate, but not enough to lose statistical significance. For the traditional RDS weights, both methods give similarly biased point estimates that are statistically different from baseline.}
\end{figure}

\begin{figure}[!ht]
    \centering
    \includegraphics[width=0.95\textwidth]{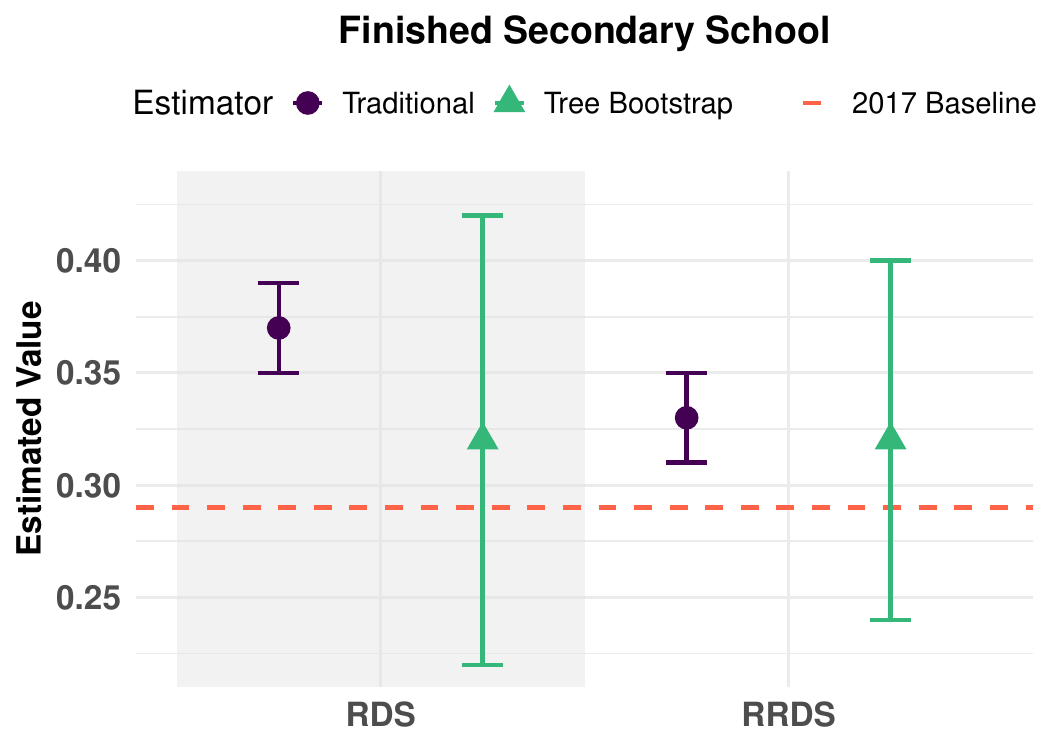}
    \caption{Proportion of respondents with secondary school education across recruitment waves. Both methods are positively biased relative to baseline, though the RRDS estimate has less uncertainty than the RDS estimate when RDS Tree Bootstrap weights are used. For traditional RDS weights, the RRDS estimate is less biased than the RDS estimate, though neither are statistically significant. }
\end{figure}

\begin{figure}[!ht]
    \centering
    \includegraphics[width=0.95\textwidth]{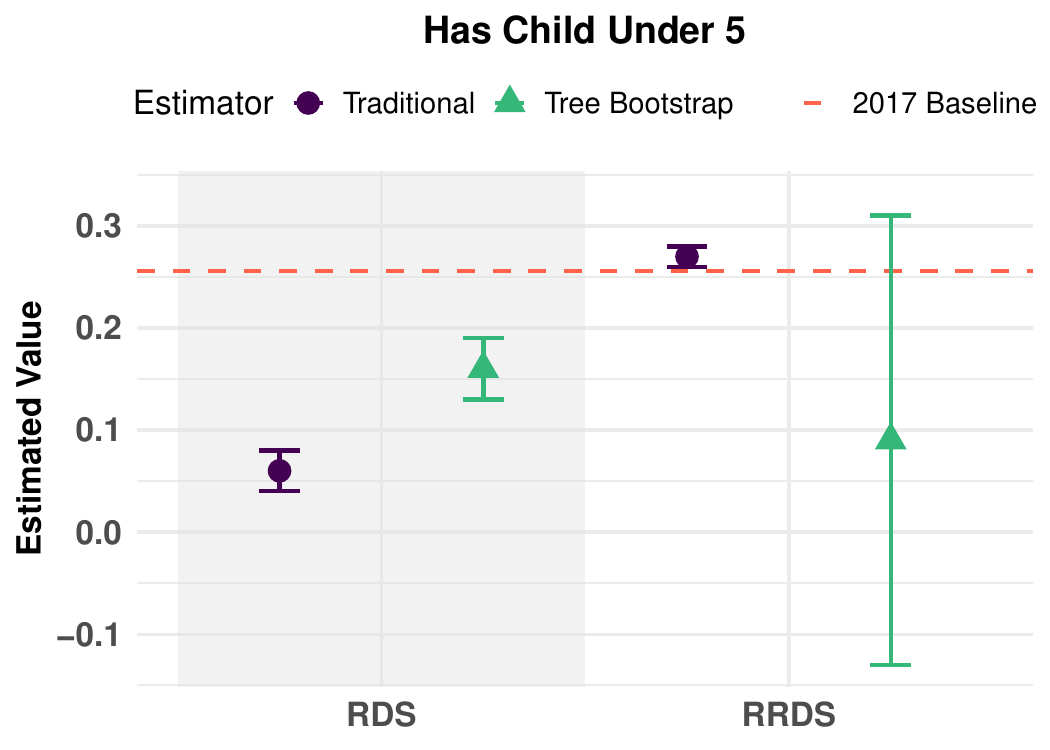}
    \caption{Proportion of respondents with children under 5 years old currently living at home across recruitment waves. This demographic characteristic shows particularly strong homophily effects in traditional RDS. We find considerable uncertainty around the RRDS estimate with RDS Tree Bootstrap weights. However, with traditional RDS weights, the RRDS estimate is unbiased and statistically significant.}
\end{figure}

\end{document}